\documentclass[12pt]{article}

\usepackage[margin=1in]{geometry}
\usepackage{setspace}

\usepackage{titling}
\usepackage{authblk}
\usepackage[dvipsnames]{xcolor}
\usepackage{amsmath}
\usepackage{tikz}
\usepackage{graphicx}
\usepackage{subcaption}
\usepackage{enumerate}
\usepackage[round]{natbib}
\usepackage{url}
\usepackage{caption}
\usepackage[inline]{enumitem}
\usepackage[normalem]{ulem}

\usepackage{bbm}
\renewcommand*{\P}{\mathbbm{P}}
\newcommand*{\E}{\mathbbm{E}}

\renewcommand*{\vec}[1]{\boldsymbol{#1}}

\newcommand*{\set}[1]{\mathrm{#1}}
\newcommand*{\col}[1]{\mathcal{#1}}

\newcommand*{\giv}{\;|\;}
\newcommand*{\mgiv}{\;\middle|\;}
\newcommand{\ind}{\perp\!\!\!\perp}
\newcommand{\notind}{\not\!\ind}
\newcommand*{\pa}{\boldsymbol{\mathrm{pa}}}

\usepackage{amsthm}
\theoremstyle{plain}
\newtheorem{lemma}{Lemma}
\newtheorem{theorem}{Theorem}

\theoremstyle{definition}

\newtheorem{condition}{Condition}

\usepackage{mathtools}
\usepackage{booktabs}

\colorlet{changed}{blue}
\colorlet{removed}{red}

\usetikzlibrary{shapes,decorations,arrows,calc,arrows.meta,fit,positioning}
\tikzset{
  -Latex,auto,node distance =1 cm and 1 cm,semithick,
  el/.style = {inner sep=2pt, align=left, sloped}
}

\begin{document}

\def\spacingset#1{\renewcommand{\baselinestretch}%
  {#1}\small\normalsize} \spacingset{1}

%%%%%%%%%%%%%%%%%%%%%%%%%%%%%%%%%%%%%%%%%%%%%%%%%%%%%%%%%%%%%%%%%%%%%%%%%%%%%%

\title{Graphical tools for detection and control of selection bias with multiple exposures and samples}

\author[1]{Patrick M.\ Schnell}
\author[1]{Eben Kenah}
\affil[1]{Division of Biostatistics, The Ohio State University College of Public Health}

\date{September 20, 2024}

\maketitle

\begin{abstract}
  Among recent developments in definitions and analysis of selection bias is the potential outcomes approach of Kenah (\textit{Epidemiology}, 2023), which allows non-parametric analysis using single-world intervention graphs, linking selection of study participants to identification of causal effects.
  Mohan \& Pearl (\textit{JASA}, 2021) provide a framework for missing data via directed acyclic graphs augmented with nodes indicating missingness for each sometimes-missing variable, which allows for analysis of more general missing data problems but cannot easily encode scenarios in which different groups of variables are observed in specific subsamples.
  We give an alternative formulation of the potential outcomes framework based on conditional separable effects and indicators for selection into subsamples.
  This is practical for problems between the single-sample scenarios considered by Kenah and the variable-wise missingness considered by Mohan \& Pearl.
  This simplifies identification conditions and admits generalizations to scenarios with multiple, potentially nested or overlapping study samples, as well as multiple or time-dependent exposures.
  We give examples of identifiability arguments for case-cohort studies, multiple or time-dependent exposures, and direct effects of selection.
\end{abstract}

\noindent%

% \spacingset{1.9}
\doublespacing

\section{Introduction}

Selection bias is a fundamental threat to valid epidemiologic inferences.
Recently, significant attention has returned to detecting and controlling selection bias using graphical tools similar to those widely used to analyze confounding.
\cite{hernan2004structural} present a structural approach via directed acyclic graphs (DAGs, \cite{greenland1999causal}) and show that selection bias occurs when (but not only when) conditioning on selection into a sample opens a non-causal path from exposure to outcome.
However, the structural approach can only detect selection bias that can occur in the absence of a causal effect of exposure on outcome \citep{greenland1977response, kenah2023potential}.
\cite{lu2022toward} also use DAGs to distinguish between failures of internal validity (collider bias, or ``type 1 selection bias'') and external validity (generalizability bias, or ``type 2 selection bias'').
\cite{sjolander2023selection} notes that a weakness of this approach is that its relationship to potential outcomes and associated exchangeability assumptions is opaque as they do not appear in the DAGs.

\cite{kenah2023potential} proposes a definition of selection bias in terms of potential outcomes which captures selection bias under both the structural definition of \cite{hernan2004structural} and the traditional difference-in-measures definition and allows for the simultaneous analysis of confounding and selection bias using single-world intervention graphs (SWIGs, \cite{richardson2013single}).
\cite{sjolander2023selection} uses counterfactual graphs \citep{shpitser2007counterfactuals} to contrast outcome-associated selection (where there is an open non-causal path from the outcome to selection) and outcome-influenced selection (where at least one open path from the outcome to selection is causal), and shows that causal effects in the selected population are identifiable under the former but not the latter.
Further, there is no information about causal effects in the full population under outcome-associated selection, but, as is shown by \cite{kenah2023potential}, under outcome-influenced sampling the causal odds ratio in the source population is identifiable.
\cite{sjolander2023selection} does not consider the impact of other sources of bias, including confounding.
\cite{hernan2004structural}, \cite{kenah2023potential}, and \cite{sjolander2023selection} consider identifiability in a single selected population or the source population from a single sample in which all variables are fully observed.

\cite{mohan2021graphical} develop the concept of an m-graph which provides for each partially missing variable $V$ a missingness indicator $H_V$ (for ``hidden'' in our notation) and a proxy variable $V^*$ defined as $V$ if $H_V=0$ and some placeholder $\eta$ if $H_V=1$.
Classification of missing data mechanisms into missing completely at random, missing not at random, and a variation of missing at random are expressible in terms of unconditional and conditional independencies among fully observed, partially observed, and fully unobserved variables, along with missingness indicators.
Recoverability (a broadening of ``identifiability'' to apply to non-causal relationships) results are also presented in terms of such independencies, which can be evaluated from m-graphs.
The m-graph approach allows for analysis of general missing data problems, and it can be adapted to situations in which all variables are either observed or unobserved simultaneously (as in selection).
However, it cannot easily encode scenarios in which different groups of variables are observed in specific subsamples.

Transportability literature (e.g., \cite{bareinboim2013general}) considers inferences made about one population from another, and often considers cases with two populations in which different sets of variables are observed, or there are slightly different causal structures (e.g., randomized versus non-randomized exposures).
One common scenario is extension of inferences from a randomized trial to a target population \citep{dahabreh2019extending}.

Our objective is to provide graphical tools to aid in detection and control of selection bias among multiple samples in which different sets of variables may be observed.
We do so by revisiting the potential outcomes approach of \cite{kenah2023potential} through the lens of separable effects \citep{robins2022interventionist, stensrud2022separable, stensrud2023conditional}, a recent development in causal mediation analysis that helps to clarify when conditioning on (possibly post-exposure) variables does not irrecoverably obscure the causal effect of exposure.
In doing so, we clarify and relax the identifiability conditions of \cite{kenah2023potential}.
Our approach also allows for analysis of selection bias in the presence of time-dependent or otherwise multiple exposures beyond the simple case of monotone censoring \citep{robins2000marginal, hernan2001marginal}.

\section{Graphical models for selection bias and missing data}

\subsection{m-graphs for missing data}

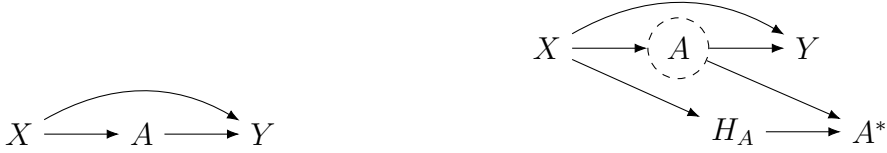
\begin{figure}
  \centering
  \begin{subfigure}[t]{0.4\textwidth}
    \begin{tikzpicture}
      \node (x) at (0, 0) {$X$};
      \node (a) [right =of x] {$A$};
      \node (y) [right =of a] {$Y$};

      \path (x) edge (a);
      \path (a) edge (y);
      \path (x) edge[bend left=30] (y);
    \end{tikzpicture}
  \end{subfigure} %
  ~
  \begin{subfigure}[t]{0.4\textwidth}
    \begin{tikzpicture}
      \node (x) at (0, 0) {$X$};
      \node (a) [right =of x, draw, dashed, circle] {$A$};
      \node (y) [right =of a] {$Y$};

      \path (x) edge (a);
      \path (a) edge (y);
      \path (x) edge[bend left=30] (y);

      \node (ha) [below right =0.5 and 0 of a] {$H_A$};
      \node (pa) [right =of ha] {$A^*$};

      \path (a) edge (pa);
      \path (ha) edge (pa);
      \path (x) edge (ha);
    \end{tikzpicture}
  \end{subfigure}
  \caption{
    A causal directed acyclic graph (DAG) under no missingness (left) and an m-graph \citep{mohan2021graphical} in which the dashed circle around $A$ indicates that it is treated as unobserved (right).
    The variable $A$ in the m-graph is ``missing at random'' given $X$ because it is independent of its missingness indicator $H_A$ given $X$.
    Removal of the $X \rightarrow H_A$ arrow would represent ``missingness completely at random'' and the addition of an $A \rightarrow H_A$  arrow would represent ``missingness not at random.''
  }
  \label{fig:m-graphs}
\end{figure}

In the m-graph approach of \cite{mohan2021graphical}, the original variable $V$ is treated as unobserved, and given a partition of the graph variables into fully observed $\vec{V}_o$, partially missing $\vec{V}_m$, fully unobserved $\vec{U}$, proxy $\vec{V}^*$, and missingness indicator $\vec{H}$ variables, the following classes of missingness mechanisms are defined:
\begin{itemize}
\item \textbf{MCAR:} $\vec{V}_m \cup \vec{V}_o \cup \vec{U} \ind \vec{H}$,
\item \textbf{v-MAR:} $\vec{V}_m \cup \vec{U} \ind \vec{H} \giv \vec{V}_o$,
\item \textbf{MNAR:}  Neither MCAR nor v-MAR.
\end{itemize}

Figure~\ref{fig:m-graphs} depicts an example of an m-graph constructed from a DAG.
A primary result of \cite{mohan2021graphical} is that, given an m-graph and an observed data distribution $\P[\vec{V}^*, \vec{V}_o, \vec{H}]$, a target quantity $\psi$ is recoverable if it can be decomposed into an ordered factorization compatible with the DAG, or a sum of such factorizations, such that every factor $\rho_j = \P[\vec{V}_{jL} \giv \vec{V}_{jR}]$ satisfies $\vec{V}_{jL} \ind (\vec{H}_{V_{jL}}, \vec{H}_{V_{jR}}) \giv \vec{V}_{jR}$.
Then each $\rho_j$ may be recovered as $\P[ \vec{V}_{jL}^* \giv \vec{V}_{jR}^*, \vec{H}_{\vec{V}_{jL}}=\vec{0}, \vec{H}_{\vec{V}_{jR}}=\vec{0} ]$.

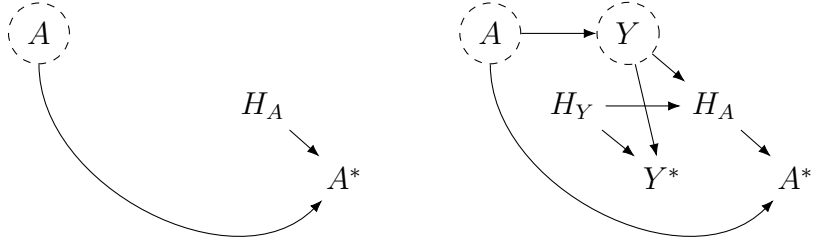
\begin{figure}
  \centering
  \begin{tikzpicture}
    \node (A2) [draw, circle, dashed] at (0, 0) {$A$};
    \node (Y2) [right =of A2] {$\phantom{Y}$};
    \node (HY2) [below right =0.5 of A2] {$\phantom{H_Y}$};
    \node (HA2) [right =of HY2] {$H_A$};
    \node (Yp2) [below right =0.5 of HY2] {$\phantom{Y^*}$};
    \node (Ap2) [right =of Yp2] {$A^*$};

    \path (A2) edge[out=-90, in=-135] (Ap2);
    \path (HA2) edge (Ap2);
    
    \node (A) [draw, circle, dashed] at (6, 0) {$A$};
    \node (Y) [draw, circle, dashed, right =of A] {$Y$};
    \node (HY) [below right =0.5 of A] {$H_Y$};
    \node (HA) [right =of HY] {$H_A$};
    \node (Yp) [below right =0.5 of HY] {$Y^*$};
    \node (Ap) [right =of Yp] {$A^*$};

    \path (A) edge (Y);
    \path (A) edge[out=-90, in=-135] (Ap);
    \path (Y) edge (HA);
    \path (Y) edge (Yp);
    \path (HY) edge (HA);
    \path (HY) edge (Yp);
    \path (HA) edge (Ap);

  \end{tikzpicture}
  \caption{The m-graphs corresponding to a study with two samples from the same population: a representative sample of exposure (left), and a case-control sample (right).}
  \label{fig:one-study-two-m-graphs}
\end{figure}

Two obstacles pose difficulties for applying m-graphs directly to the type of problem we consider.
First, we wish to work primarily with single world intervention graphs (or another causal graph that represents potential outcomes explicitly) to facilitate the analysis of causal selection problems in terms of potential outcomes.
The interpretation and use of missingness and proxy variables for potential outcomes is unclear because, even without missingness as usually conceptualized, the observability of potential outcomes is determined by the actual exposure.
Second, m-graphs excel at encoding unconstrained independencies between missingness and other variables, but they struggle to encode more complex relationships.
For example, consider a study in which two non-nested samples are drawn from a population: one is a representative sample in which an exposure $A$ is measured, and the other is a case-control sample which is dependent on the outcome $Y$ but independent of $A$ conditional on $Y$.
From the first sample, we can recover the distribution $\P[A=a]$.
From the second sample, we can recover the conditional distributions $\P[A=a \giv Y=y]$.
These results can be combined to recover the risk ratio $\P[Y=1 \giv A=1] / \P[Y=1 \giv A=0]$.
Such a scenario requires two m-graphs to properly encode (Figure~\ref{fig:one-study-two-m-graphs}), and it may be difficult to keep track of cross-sample assumptions such as independence of selection into the two samples.
In Section~\ref{sec:case-cohort} we show that our approach is straightforward to apply even for a more complex version of this example.

\subsection{Selection bias with single samples}

\cite{kenah2023potential} proposes a definition of selection bias in terms of potential outcomes which captures selection bias under both the structural definition and the difference-in-measures definition and allows for the simultaneous analysis of confounding and selection bias using SWIGs.
This potential outcomes approach to selection bias primarily considers identification of a causal effect in a population from a single sample drawn from it in which all relevant variables are fully observed.
In this apprach, there is no unmeasured selection bias for the causal effect of $A$ on $Y$ if and only if at least one of the following conditions holds:

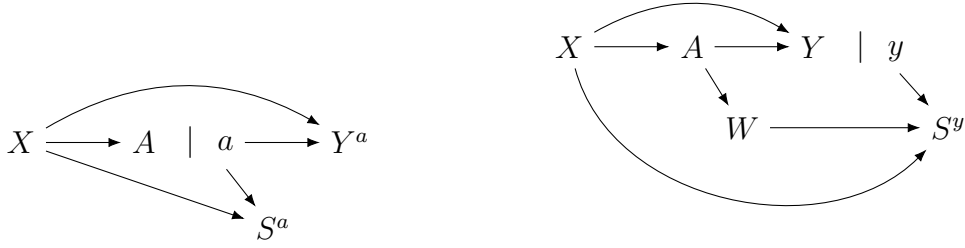
\begin{figure}
  \centering
  \begin{subfigure}[t]{0.4\textwidth}
    \begin{tikzpicture}
      \node (x) at (0, 0) {$X$};
      \node (A) [right =of x] {$A$};
      \node (a) [right =0 of A] {$\giv \;\, a$};
      \node (y) [right =of a] {$Y^a$};

      \path (x) edge (A);
      \path (a) edge (y);
      \path (x) edge[bend left=30] (y);

      \node (s) [below right =0.5 and 0 of a] {$S^a$};

      \path (x) edge (s);
      \path (a) edge (s);
    \end{tikzpicture}
  \end{subfigure} %
  ~\quad
  \begin{subfigure}[t]{0.4\textwidth}
    \begin{tikzpicture}
      \node (x) at (0, 0) {$X$};
      \node (a) [right =of x] {$A$};
      \node (Y) [right =of a] {$Y$};
      \node (y) [right =0 of Y] {$\giv \;\, y$};

      \path (x) edge (a);
      \path (a) edge (Y);
      \path (x) edge[bend left=30] (Y);

      \node (w) [below right =0.5 and 0 of a] {$W$};
      \node (s) [right =2 of w] {$S^y$};

      \path (x) edge[out=-75, in=-135] (s);
      \path (y) edge (s);
      \path (a) edge (w);
      \path (w) edge (s);
    \end{tikzpicture}
  \end{subfigure}
  \caption{
    (Left) A single world intervention graph (SWIG) with a selection variable $S^a$ indicating whether all (versus none) of the other variables on the graph are observed.
    Kenah's analytic cohort condition for this SWIG requires $S^a \ind Y^a \giv (A, X)$ and $Y^a \ind A \giv X$.
    (Right) A SWIG of the type used to evaluate Kenah's analytic case-control condition.
    The condition requires $S^y \ind A \giv (Y, X, W)$ as well as $Y^a \ind A \giv X$ and $W^a \ind Y^a \giv (A, X)$, the latter two not readable from this SWIG.
  }
  \label{fig:swig-kenah}
\end{figure}

\begin{condition}[Analytic cohort condition \citep{kenah2023potential}]
  If we intervene to set exposure $A=a$, we have
  \begin{align}
    A \ind Y^a &\giv \vec{X}, \\
    S^a \ind Y^a &\giv (A, \vec{X}),
  \end{align}
  where $\vec{X}$ is a possibly empty set of measured non-descendants of $a$.
\end{condition}

\begin{condition}[Analytic case-control condition \citep{kenah2023potential}]
  If we intervene to set exposure $A=a$, we have
  \begin{align}
    A \ind Y^a &\giv \vec{X}, \\
    \vec{W}^a \ind Y^a &\giv (A, \vec{X}), \label{eq:case-control-w}
  \end{align}
  where $\vec{X}$ is a set of measured non-descendants of $a$ and $\vec{W}$ is a set of measured non-descendants of $y$.
  Further, if we intervene to set outcome $Y=y$ (but not exposure), we have
  \begin{equation}
    S^y \ind A \giv (Y, \vec{X}, \vec{W}).
  \end{equation}
\end{condition}

Figure~\ref{fig:swig-kenah} depicts example SWIGs satisfying the analytic cohort and case-control conditions.
If the analytic cohort condition is satisfied, then $\P[Y^a=y \giv \vec{X}] = \P[Y^a=y \giv A=a, S^a=1, \vec{X}] = \P[Y=y \giv A=a, S=1, \vec{X}]$, and any measure of causal effect based on these conditional distributions of potential outcomes is the same in the population and the sample up to random variation.
If the analytic case-control condition is satisfied, then the conditional causal odds ratio given $\vec{X}$ in the population is equal to the conditional exposure odds ratio given $(\vec{X},\vec{W})$ in the sample up to random variation:
\begin{align}
  &\frac{
    \P[Y^1=1 \giv \vec{X}=\vec{x}] / \P[Y^1=0 \giv \vec{X}=\vec{x}]
    }{
    \P[Y^0=1 \giv \vec{X}=\vec{x}] / \P[Y^0=0 \giv \vec{X}=\vec{x}]
    } \\
  &\quad=
    \frac{
    \P[A=1 \giv Y=1, \vec{W}=\vec{w}, \vec{X}=\vec{x}, S=1] / \P[A=0 \giv Y=1, \vec{W}=\vec{w}, \vec{X}=\vec{x}, S=1]
    }{
    \P[A=1 \giv Y=0, \vec{W}=\vec{w}, \vec{X}=\vec{x}, S=1] / \P[A=0 \giv Y=0, \vec{W}=\vec{w}, \vec{X}=\vec{x}, S=1]
    }.\nonumber
\end{align}
The key to identification is that the factual outcome odds ratio in terms of $\P[Y=y \giv A=a, \vec{X}=\vec{x}, \vec{W}=\vec{w}]$ is equal to the exposure odds ratio in terms of $\P[A=a \giv Y=y, \vec{X}=\vec{x}, \vec{W}=\vec{w}]$ in the population, and the latter is equal to $\P[A=a \giv Y=y, \vec{X}=\vec{x}, \vec{W}=\vec{w}, S^y=1]$ because $A \ind S^y \giv (Y, X, W)$ and $S^y$ may be replaced by $S$ due to consistency.
While the analytic cohort condition allows estimation of within-sample causal effects, the analytic case-control condition does not.
When only the latter holds, the observed odds ratio in the sample is not generally equal to the corresponding causal odds ratio, even conditional on $(\vec{W},\vec{X})$.
This reflects a fundamental difference between cohort and case-control studies.

We aim to simplify and harmonize the above conditions, and, in doing so, allow extension to multiple samples.
In particular, the case-control condition above has two weaknesses: the requirement to evaluate the independence conditions on two separate SWIGs $\col{G}^a$ and $\col{G}^y$, and an opaque independence condition \eqref{eq:case-control-w}.
Further, we will see that the non-descent conditions of both the cohort and case-control conditions can be slightly relaxed.

\section{Intermediate tools for multiple samples}

\subsection{Within-sample identification of causal effects via separability}

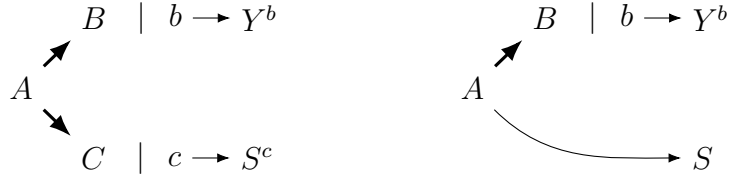
\begin{figure}
  \centering
  \begin{tikzpicture}
    \node (A1) at (0, 0) {$A$};
    \node (B1) [above right =0.5 of A1] {$B$};
    \node (b1) [right =0 of B1] {$\giv \;\, b$};
    \node (C1) [below right =0.5 of A1] {$C$};
    \node (c1) [right =0 of C1] {$\giv \;\, c$};
    \node (Y1) [right =0.5 of b1] {$Y^b$};
    \node (S1) [right =0.5 of c1] {$S^c$};

    \path (A1) edge[very thick] (B1);
    \path (A1) edge[very thick] (C1);
    \path (b1) edge (Y1);
    \path (c1) edge (S1);
    
    \node (A2) at (6, 0) {$A$};
    \node (B2) [above right =0.5 of A2] {$B$};
    \node (b2) [right =0 of B2] {$\giv \;\, b$};
    \node (C2) [below right =0.5 of A2] {$\phantom{C}$};
    \node (c2) [right =0 of C2] {$\phantom{\giv \;\, c}$};
    \node (Y2) [right =0.5 of b2] {$Y^b$};
    \node (S2) [right =0.5 of c2] {$S$};

    \path (A2) edge[very thick] (B2);
    \path (A2) edge[out=-45, in=180] (S2);
    \path (b2) edge (Y2);
  \end{tikzpicture}
  \caption{
    (Left) The separated SWIG $\col{G}_{\textrm{ex}}^{bc}$ for separated intervention on exposure constructed from the same underlying DAG as the SWIG $\col{G}^a$ in Figure~\ref{fig:swig-kenah} but without a confounder $X$.
    (Right) The SWIG $\col{G}_{\textrm{ex}}^b$ for intervening on $B$ only.
  }
  \label{fig:separable-cohort}
\end{figure}

\begin{figure}
  \centering
  \begin{subfigure}[t]{0.4\textwidth}
    \begin{tikzpicture}
      \node (X) at (0, 0) {$X$};
      \node (A) [right =0.5 of X] {$A$};
      \node (B) [right=0.5 of A] {$B$};
      \node (b) [right =0 of B] {$\giv \;\, b$};
      \node (C) [below right=0.5 of A] {$C$};
      \node (c) [right =0 of C] {$\giv \;\, c$};
      \node (Y) [right =0.5 of b] {$Y^{b}$};

      \path (A) edge[very thick] (B);
      \path (A) edge[very thick] (C);
      
      \path (X) edge (A);
      \path (b) edge (Y);
      \path (X) edge[bend left=30] (Y);
      
      \node (W) [below right =0.5 and 0 of c] {$W^c$};
      \node (S) [right =1.5 of W] {$S^{b,c}$};
      
      \path (X) edge[out=-75, in=-135] (S);
      \path (Y) edge (S);
      \path (c) edge (W);
      \path (W) edge (S);
    \end{tikzpicture}
  \end{subfigure} %
  ~\quad
  \begin{subfigure}[t]{0.4\textwidth}
    \begin{tikzpicture}
      \node (X) at (0, 0) {$X$};
      \node (A) [right =0.5 of X] {$A$};
      \node (b) [right =0 of A] {$\giv \;\, b$};
      \node (C) [below right=0.5 of A] {$\phantom{C}$};
      \node (c) [right =0 of C] {$\phantom{\giv \;\, c}$};
      \node (Y) [right =1 of b] {$Y^{b}$};

      \path (A) edge (W);
      
      \path (X) edge (A);
      \path (b) edge (Y);
      \path (X) edge[bend left=30] (Y);
      
      \node (W) [below right =0.5 and 0 of c] {$W^{\phantom{c}}$};
      \node (S) [right =1.5 of W] {$S^{b\phantom{,c}}$};
      
      \path (X) edge[out=-75, in=-135] (S);
      \path (Y) edge (S);
      \path (W) edge (S);
    \end{tikzpicture}
  \end{subfigure}
  \caption{
    (Left) The separated SWIG $\col{G}_{\textrm{ex}}^{b,c}$ for separated intervention on exposure constructed from the same underlying case-control DAG as the SWIG $\col{G}^y$ for intervening on outcome in Figure~\ref{fig:swig-kenah}.
    (Right) The abbreviated separated SWIG $\col{G}_{\textrm{ex}}^b$ for intervening only on the exposure component directly affecting the outcome.
  }
  \label{fig:separable-case-control}
\end{figure}
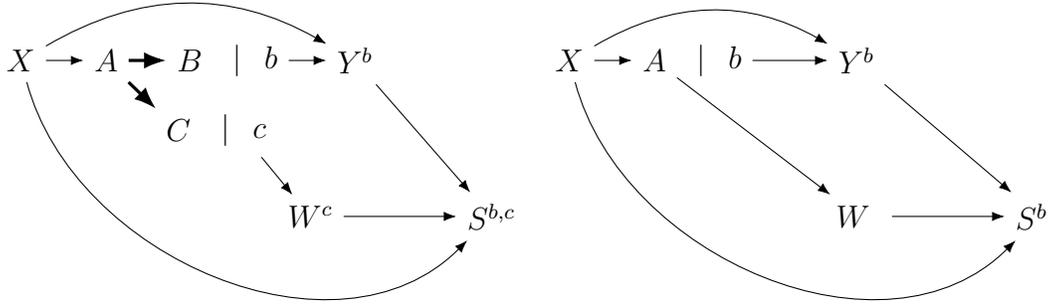

The fact that within-sample causal effects can be identified under the cohort but not case-control conditions is clarified by concepts from causal mediation analysis.
Intuitively, we can identify the within-sample causal effect if we can decompose the exposure $A$ into two components $B$ and $C$ such that
\begin{enumerate*}[label=(\roman*)]
\item \label{it:effect-outcome} $B$ has the same effect on the outcome as the original exposure, but
\item \label{it:effect-selection} $B$ does not affect selection, and
\item that effect is identifiable from within the selected sample.
\end{enumerate*}
In the available data, we have the deterministic relation $A \equiv B \equiv C$ but can imagine hypothetical interventions setting $B$ and $C$ to different values.
Conditions \ref{it:effect-outcome} and \ref{it:effect-selection} respectively can be expressed formally as the following:
\begin{condition}[Totality]
  \label{cnd:totality}
  All causal paths from $A$ to $Y$ pass through $B$.
\end{condition}
\begin{condition}[Cohort isolation]
  \label{cnd:cohort-isolation}
  There are no causal paths from $B$ to $S$.
\end{condition}

Condition~\ref{cnd:totality} means that $Y^{b=\beta} = Y^{a=\beta}$ so that if we can identify, e.g., $\P[Y^{b=\beta} = y \giv A=\beta, S=1]$ then we can equate it to $\P[Y^{a=\beta} = y \giv A=\beta, S=1]$.
Condition~\ref{cnd:cohort-isolation} is equivalent to Definition~1 of \cite{stensrud2023conditional}, which they call ``$A_Y$ partial isolation'' and use to define treatment effects on outcomes conditional on post-treatment events, in our case selection.
The isolation condition holds automatically when selection precedes exposure, as in a clinical trial.
In an observational cohort study, it can hold even when there is a causal effect of exposure $A$ on selection, as long as it passes through $C$ rather than $B$.
In case-control studies, only the following weaker condition holds:
\begin{condition}[Case-control isolation]
  \label{cnd:case-control-isolation}
  All causal paths from $B$ to $S$ pass through $Y$.
\end{condition}

The above conditions are expressible graphically.
From the original DAG $\col{G}$, we construct an expanded DAG $\col{G}_{\textrm{ex}}$ by adding $B$ and $C$ as deterministic child nodes of $A$ which have no other parents, and move the origin of each outgoing arrow from $A$ to one of $B$ and $C$.
All arrows leaving $B$ and none leaving $C$ should be part of a causal path from $A$ to $Y$ (satisfying the totality condition).
We then construct the $(b,c)$-SWIG $\col{G}_{\textrm{ex}}^{bc}$ by the usual node-splitting operation for $B$ and $C$.
Such a SWIG always exists and is unique if we allow one or both of $B$ and $C$ to have no children.
The SWIGs in Figures~\ref{fig:separable-cohort} (left) and \ref{fig:separable-case-control} (left) are examples of the resulting separated SWIGs.
Totality can be evaluated by checking that $Y^{b,c}$ is not a descendant of $c$ (which is automatic by proper construction), cohort isolation by checking that $S^{b,c}$ is not a descendant of $b$, and case-control isolation by checking that any directed path from $b$ to $S^{b,c}$ passes through $Y^{b,c}$.
For the purpose of analyzing selection bias, it suffices to use the SWIG $\col{G}_{\textrm{ex}}^{b}$ which represents intervention only on $B$ (e.g., Figure~\ref{fig:separable-cohort}, right).
This may be abbreviated by merging $A$, $B$, and $C$ so that we have a split node $(A \giv b)$ in which arrows leaving $C$ in $\col{G}_{\textrm{ex}}$ are shown leaving $A$, as in Figure~\ref{fig:separable-case-control} (right).
\begin{figure}
  \centering
  \begin{tikzpicture}
    \node (A) at (0, 0) {$A$};
    \node (b) [right =0 of A] {$\giv \;\, b$};
    \node (M) [right =of b] {$M^{b}$};
    \node (Y) [right =of M] {$Y^b$};
    
    \path (b) edge (M);
    \path (M) edge (Y);
    
    \node (S) [below right =of Y] {$S^{b}$};
    
    \path (M) edge (S);
    \path (Y) edge[dashed] (S);
  \end{tikzpicture}
  \caption{
    An abbreviated SWIG $\col{G}_{\textrm{ex}}^b$ in which the effect of exposure on selection is not separable with respect to the outcome, regardless of whether the dashed $Y^b \rightarrow S^b$ arrow is present.
  }
  \label{fig:nonseparable}
\end{figure}
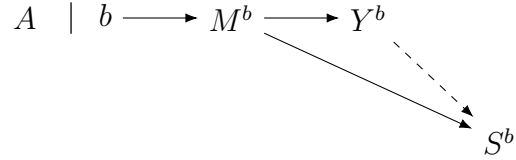
Figure~\ref{fig:nonseparable} depicts a $b$-SWIG in which the isolation conditions fail due to the presence of a confounder $M^b$ of $S^b$ and $Y^b$ that itself is a descendant of $b$.
We discuss with examples in Appendix~\ref{app:a-swig-b-swig} the additional assumptions implied by a $b$-SWIG compared to the original SWIG.
Specifically, if some set of variables $\vec{W}$ consists of descendants of $B$ but not $C$, and some set of variables $\vec{X}$ consists of descendants of $C$ but not $B$, then the $b$-SWIG makes the assumptions that $\vec{W}^{a=1} \ind \vec{X}^{a=0}$ and $\vec{W}^{a=0} \ind \vec{X}^{a=1}$.
These assumptions (and possibly many more) are already made by non-parametric structural equation models with independent errors (NPSEM-IE) associated with the underlying DAG $\col{G}$, but are not made by the finest fully randomized causally interpretable structured tree graph (FFRCISTG) associated with the $a$-SWIG $\col{G}^a$ \citep{robins1986new, shpitser2022multivariate}.
They are, however, made by the FFRCISTG associated with the $(b,c)$-SWIG $\col{G}_{\text{ex}}^{bc}$ under the determinism assumption $a \equiv B^a \equiv C^a$, and are not cross-world if $B$ and $C$ correspond to real components of $A$ that can be intervened on separately.

Identifiability of the within-sample causal effects under the cohort condition (e.g., Figure~\ref{fig:separable-cohort}, right) but not the case-control condition (e.g., Figure~\ref{fig:separable-case-control}, right) is a consequence of the fact that on $\col{G}_{\textrm{ex}}^b$ the factual selection variable $S$ appears in the former but only the potential selection variable $S^b$ appears in the latter.
Cohort isolation is necessary for the analytic cohort condition to hold (Appendix~\ref{app:cohort-isolation-necessary}), and thus $S^b=S$ under the latter.
Thus if we can meet the analytic cohort condition, we can identify the effect of the outcome-affecting exposure component in the actual sample ($B$ on $Y$ given $S=1$), which is equal to the effect of the full exposure in the actual sample ($A$ on $Y$ given $S=1$).
However, only case-control isolation is necessary for the analytic case-control condition (Appendix~\ref{app:case-control-isolation-necessary}), so selection remains affected by the outcome-affecting exposure component and restriction to the sample is not equivalent to conditioning on $S^b=1$.

Under the usual conditions of consistency, positivity, and conditional exchangeability \cite{stensrud2023conditional} give additional dismissibility conditions sufficient to identify $\E[Y^{bc} \giv D^{bc}=1]$ for some post-treatment variable $D$, even when $b \neq c$.
These dismissibility conditions yield the following cohort condition with relaxed non-descent conditions expressed in terms of our $b$-SWIG approach.
\begin{condition}[Separable cohort condition]
  \label{cnd:sep-cohort}
  If we intervene to set $B = b$ on $\col{G}_{\textrm{ex}}^b$, exchangeability holds and potential selection $S^b$ is independent of the potential outcome $Y^b$ conditional on $A$ and a set $\vec{K}$ of non-descendants of $b$:
  \begin{align}
    A \ind Y^b &\giv \vec{K}, \\
    S^b \ind Y^b &\giv (A, \vec{K}).
  \end{align}
  The above pair of independencies is equivalent to $(A, S^b) \ind Y^b \giv \vec{K}$.
\end{condition}
Note that under this cohort condition, $S$ will not be a descendant of $b$, in which case by consistency $S^b = S$.
If we further require that $\vec{K}$ contains no descendants of $A$ on $\col{G}$, it is equivalent to the analytic cohort condition (Appendix~\ref{app:cohort-equivalence}).

\subsection{Clarification of the analytic case-control condition}

The case-control isolation condition clarifies the function of the secondary independence condition \eqref{eq:case-control-w} that $\vec{W}^a \ind Y^a \giv (A, \vec{X})$ in the analytic case-control condition.
The analytic case-control condition includes the independence $S^y \ind A \giv (Y, \vec{X}, \vec{W})$, which conditions on the observed variable $Y$ not appearing on our separated SWIGs.
The following lemma, proved in Appendix~\ref{app:observable-conditioning}, connects a conditional independence relationship between observed variables $S$ and $Y$ to one between the potential variables $S^b$ and $Y^b$ appearing on the separated SWIG:
\begin{lemma}[Observable conditioning]
  \label{lem:observable-conditioning}
  Under Condition~\ref{cnd:case-control-isolation} (case-control isolation) that all causal paths from $B$ to $S$ pass through $Y$, for any (possibly empty) set of variables $\vec{K}$ not containing descendants of $b$, we have $S^b \ind A \giv (Y^b, \vec{K})$ if and only if $S \ind A \giv (Y, \vec{K})$.
\end{lemma}
Then, under case-control isolation, if we can find a suitable set of variables $\vec{K}$ not containing descendants of $b$ so that $S^b \ind A \giv (Y^b, \vec{K})$, we can equate the distributions $[A \giv Y, \vec{K}] \sim [A \giv Y, \vec{K}, S=1]$.
The exposure odds ratio is equal to the (observed, not necessarily causal) outcome odds ratio, so if exchangeability holds given $\vec{K}$ we can identify the population causal odds ratio.
We collect the necessary conditions into the following analytic case-control condition with relaxed non-descent conditions expressed in terms of our $b$-SWIG approach.
\begin{condition}[Separable case-control condition]
  \label{cnd:sep-case-control}
  The case-control isolation condition holds.
  Further, if we intervene to set $B = b$ on $\col{G}_{\textrm{ex}}^b$, exchangeability holds and potential selection $S^b$ is independent of $A$ conditional on the potential outcome $Y^b$ and a set $\vec{K}$ of non-descendants of $b$:
  \begin{align}
    A \ind Y^b &\giv \vec{K}, \label{eq:sep-case-control-exch} \\
    S^b \ind A &\giv Y^b, \vec{K}.
  \end{align}
  The above pair of independencies is equivalent to $(Y^b, S^b) \ind A \giv \vec{K}$.
\end{condition}
If we decompose $\vec{K}$ into $(\vec{X}, \vec{W})$ so that $\vec{X}$ contains no descendants of $A$ and $\vec{W}$ is not conditioned on in \eqref{eq:sep-case-control-exch}, this condition is equivalent to the analytic case-control condition (Appendix~\ref{app:case-control-equivalence}).
Given the other components of the analytic case-control condition, the independence $\vec{W}^a \ind Y^a \giv (A, \vec{X})$ implies case-control isolation because it ensures that $\vec{W}$ does not lie on the causal path between $A$ and $Y$.

Conditioning on the potential outcome $Y^b$ in the separable case-control condition is useful mathematically but may be alarming philosophically because we cannot condition on potential outcomes in practice.
However, Lemma~\ref{lem:observable-conditioning} allays this concern by making it equivalent to $A \ind S \giv (Y, \vec{K})$.
Alternatively, the compact version of the independencies, $(Y^b, S^b) \ind A \giv \vec{K}$, avoids explicitly conditioning on the potential outcome.

\subsection{Identifiability in one sample using others}

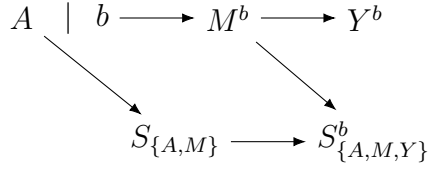
\begin{figure}
  \centering
  \begin{tikzpicture}
    \node (A) at (0, 0) {$A$};
    \node (b) [right =0 of A] {$\giv \;\, b$};
    \node (M) [right =of b] {$M^{b}$};
    \node (Y) [right =of M] {$Y^b$};
    
    \path (b) edge (M);
    \path (M) edge (Y);

    \node (S_AM) [below right =of A] {$S_{\{A,M\}}$};
    \node (S_AMY) [right =of S_AM] {$S_{\{A,M,Y\}}^{b}$};

    \path (A) edge (S_AM);
    \path (M) edge (S_AMY);
    \path (S_AM) edge (S_AMY);
  \end{tikzpicture}
  \caption{
    An example in which the analytic and separable cohort conditions fail but the causal effect is identifiable in the full population and $\set{S}_{\{A,M\}}$, though not in $\set{S}_{\{A,M,Y\}}$.
  }
  \label{fig:nonseparable-sequential}
\end{figure}

For a set of variables $\vec{V}$, we will write $S_{\{\vec{V}\}}=1$ to indicate that $\vec{V}$ is observed, without implying anything about variables not included in $\vec{V}$.
However, $S_{\{\vec{V}\}}=0$ will not imply anything about the observation of $\vec{V}$ (in particular, it does not imply that $\vec{V}$ is unobserved).
Intuitively, $S_{\{\vec{V}\}}=1$ indicates selection into some sample for which $\vec{V}$ is observed, and where other variables may or may not be observed, while $\vec{V}$ may or may not be observed outside of that sample.

In Figure~\ref{fig:nonseparable}, the effect of $A$ on $Y$ within the sample is not identifiable even in the absence of the dashed $Y^b \rightarrow S^b$ arrow.
However, it is identifiable in the population given external information about the conditional distribution $[M \giv A=a]$ for each $a$.
Consider the scenario in Figure~\ref{fig:nonseparable-sequential} where we observe $(A,M,Y)$ in some sample $\set{S}_{\{A,M,Y\}}$, selection into which depends on $M$, and $(A,M)$ in some sample $\set{S}_{\{A,M\}}$, selection into which only depends on $A$.
We also require $\set{S}_{\{A,M,Y\}} \subseteq \set{S}_{\{A,M\}}$.
We then have, via the g-formula for a joint intervention on $A$ and $\set{S}_{\{A,M,Y\}}$,
\begin{equation}
  \begin{aligned}
    \E\left[ Y^{a} \giv S_{\{A,M\}}=1 \right]
    &= \E\left[ Y^{a, s_{\{A,M,Y\}}=1} \giv S_{\{A,M\}}=1 \right], \\
    &= \sum_m \E\left[ Y \mgiv S_{\{A,M,Y\}}=1, M=m, A=a, S_{\{A,M\}}=1 \right] \\
    &\quad\quad\quad \times \P\left[M=m \giv A=a, S_{\{A,M\}}=1\right].
  \end{aligned}
\end{equation}
We can generalize the causal effect in $\set{S}_{\{A,M\}}$ to the full population because $S_{\{A,M\}}$ satisfies the separable cohort condition unconditionally.
This argument is an application of the following more general theorem for using information from multiple samples to identify causal effects.

\begin{theorem}[Identification between samples]
  \label{thm:identification-between-samples}
  A functional $\psi\left(\vec{\theta} \mgiv S_0 = 1\right)$ of the distribution of $\left[ \vec{V} \giv S_0=1 \right]$ is identifiable if the components of the vector $\vec{\theta}$ can be written as
  \begin{equation}
    \theta_j\left( \vec{V}_{jL} \mgiv \vec{V}_{jR}, S_0=1 \right),
  \end{equation}
  where
  \begin{equation}
    \label{eq:theta-decomp}
    \theta_j\left( \vec{V}_{jL} \mgiv \vec{V}_{jR}, S_j=1 \right)
  \end{equation}
  is identifiable and
  \begin{align}
    S_0 \ind \vec{V}_{jL} &\giv (\vec{V}_{jR}, S_j=1), \label{eq:0-giv-j} \\
    S_j \ind \vec{V}_{jL} &\giv (\vec{V}_{jR}, S_0=1). \label{eq:j-giv-0}
  \end{align}
  We allow $\vec{V}$, $\vec{V}_{jL}$, and $\vec{V}_{jR}$ to contain both observed and potential variables on a SWIG, but the selection variables $S_0$ and $S_j$ must be observed (i.e., not $S_0^b$ or $S_j^b$).
\end{theorem}

The proof is straightforward: $\theta_j\left( \vec{V}_{jL} \mgiv \vec{V}_{jR}, S_j=1 \right)$ is assumed identifiable by \eqref{eq:theta-decomp}, and is equal to $\theta_j\left( \vec{V}_{jL} \mgiv \vec{V}_{jR}, S_j=1, S_0=1 \right)$ by \eqref{eq:0-giv-j}, which is equal to $\theta_j\left( \vec{V}_{jL} \mgiv \vec{V}_{jR}, S_0=1 \right)$ by \eqref{eq:j-giv-0}.
Note that \eqref{eq:0-giv-j} is automatic if $S_j = 1$ implies $S_0=1$ (i.e., $\set{S}_j \subseteq \set{S}_0$) and vice versa for \eqref{eq:j-giv-0}.
We can also apply the theorem to identification in the source population of the SWIG by taking \eqref{eq:0-giv-j} as automatic because every sample is a subset of the source population, and by dropping the conditioning on $S_0$ in \eqref{eq:j-giv-0}.

\section{Examples}

Section~\ref{sec:case-cohort} gives an example of a case-cohort study in which the risks $\P[Y^a=1]$ are identifiable in the full population by combining multiple sub-samples from a single cohort, even though $A$ and $Y$ are observed in the same individuals only within a sample dependent on outcome.
Section~\ref{sec:time-dependent-exposures} illustrates sequential application of the graph transformations to multiple or time-dependent exposures, including the necessity of applying those transformations to exposures in reverse topological order.
Section~\ref{sec:covid} gives a longitudinal infectious disease surveillance example in which the causal effect of a selection variable is of interest and the necessary conditioning set is different among the selected and unselected.

\subsection{Causal risks from a case-cohort study}
\label{sec:case-cohort}

\begin{figure}
  \centering
  \begin{tikzpicture}[scale=1]
    \node (a) at (0,0) {$A$};
    \node (b) [right =0 of a] {$\giv b$};
    \node (y) at (4, 0) {$Y^b$};
    \node (s0) at (1,-1.5) {$S_{\emptyset}$};
    \node (sa) at (2, -3) {$S_{\{A\}}$};
    \node (sy) at (3, -1.5) {$S_{\{Y\}}$};
    \node (say) at (5, -1.5) {$S_{\{A,Y\}}^b$};

    \path (b) edge (y);
    \path (a) edge (s0);
    \path (s0) edge (sa);
    \path (s0) edge (sy);
    \path (y) edge (say);
    \path (sy) edge (say);
  \end{tikzpicture}
  \caption{
    An abbreviated separated SWIG representing a case-cohort study.
    Note that $Y$ appears in the subscript of the selection variables rather than $Y^b$ because if an actual (not potential) selection variable is $1$ then the actual (not potential) outcome $Y$ is observed.
  }
  \label{fig:case-cohort}
\end{figure}
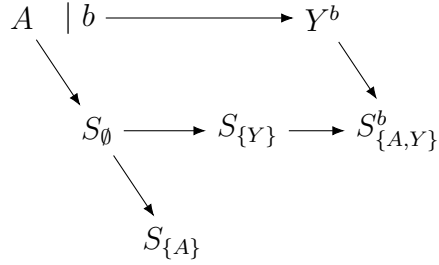

Figure~\ref{fig:case-cohort} represents a case-cohort study \citep{prentice1986case} of the following design.
A base cohort $\set{S}_{\emptyset}$ is selected in a manner that is affected by exposure, but includes both exposed and unexposed individuals.
Due to the relatively high cost of evaluating this exposure, it is only measured in a simple random sample $\set{S}_{\{A\}} \subseteq \set{S}_\emptyset$ drawn from the base cohort.
Disease status, which is comparably inexpensive to evaluate, is measured in a larger simple random sample $\set{S}_{\{Y\}} \subseteq \set{S}_\emptyset$, which could reasonably be the entire base cohort.
Finally exposure is also measured in a case sample $\set{S}_{\{A,Y\}} \subseteq \set{S}_{\{Y\}}$ which is a simple random sample (possibly all) of $\set{S}_{\{Y\}}$ with $Y=1$.

We can identify the exposure prevalence among the cases in $\set{S}_{\left\{A,Y\right\}}$, $\P\left[A=1 \mgiv Y=1, S_{\left\{A,Y\right\}}=1\right]$---and its complement---because $A$ and $Y$ are observed there.
Then, because $S_{\left\{A,Y\right\}}^b$ is d-separated from $b$ given $Y^b$ and $S_{\emptyset}$, the Lemma~\ref{lem:observable-conditioning} allows us to conclude that $S_{\left\{A,Y\right\}} \ind A \giv (Y, S_{\emptyset})$ by reading $S_{\left\{A,Y\right\}}^b \ind A \giv (Y^b, S_{\emptyset})$ from the SWIG.
By Theorem~\ref{thm:identification-between-samples} and the fact that $\set{S}_{\left\{A,Y\right\}} \subseteq \set{S}_\emptyset$, we can identify the exposure prevalence among the cases in $\set{S}_\emptyset$.

We can identify the exposure prevalence in $\set{S}_{\left\{A\right\}}$, $\P\left[A=1 \mgiv S_{\left\{A\right\}}=1\right]$, because $A$ is observed there.
Because $\set{S}_{\left\{A\right\}} \subseteq \set{S}_\emptyset$ and $S_{\left\{A\right\}} \ind A \giv S_{\emptyset}$, Theorem~\ref{thm:identification-between-samples} allows us to identify $\P\left[A=1 \mgiv S_{\emptyset}=1\right]$.
Similarly, we can identify $\P\left[Y=1 \mgiv S_{\emptyset}=1 \right] = \P\left[Y=1 \mgiv S_{\{Y\}}=1 \right]$ because $\set{S}_{\{Y\}} \subseteq \set{S}_{\emptyset}$ and $S_{\{Y\}} \ind Y \giv S_{\emptyset}$.
Then via Bayes' Theorem, consistency, and exchangeability (i.e., $Y^a \ind A \giv S_\emptyset$), we have the causal risk in $\set{S}_\emptyset$:
\begin{equation}
  \begin{aligned}
    \frac{
      \P\left[ A=a \mgiv Y=1, S_\emptyset=1 \right]
      \P\left[ Y=1 \mgiv S_{\emptyset}=1 \right]
    }{
      \P\left[ A=a \mgiv S_\emptyset=1 \right]
    }
    &=
    \P\left[Y=1 \mgiv A=a, S_\emptyset=1\right], \\
    &=
    \P\left[Y^a=1 \mgiv A=a, S_\emptyset=1\right], \\
    &=
    \P\left[Y^a=1 \mgiv S_\emptyset=1\right].
  \end{aligned}
\end{equation}
Because $Y^b \ind S_\emptyset$, this causal risk can be generalized to the source population of the SWIG by a final application of Theorem~\ref{thm:identification-between-samples}.

Though usually of limited scientific interest, Theorem~\ref{thm:identification-between-samples} allows us to also transport the causal risks from $\set{S}_\emptyset$ to $\set{S}_{\{A\}}$ or $\set{S}_{\{Y\}}$ because $\set{S}_{\{A\}} \subseteq \set{S}_\emptyset$ and $Y^b \ind S_{\{A\}} \giv S_\emptyset$---and similarly for $\set{S}_{\{Y\}}$.
Although we can identify the causal risks in the source population and observe both exposure and outcome in $\set{S}_{\{A,Y\}}$, we cannot identify the causal risks in the latter (which are often of no scientific interest).

\subsection{Multiple and time-dependent exposures}
\label{sec:time-dependent-exposures}

\begin{figure}
  \centering
  \begin{tikzpicture}[scale=1]
    \node (a1) at (0, 0) {$A_1$};
    \node (b1) [right =0 of a1] {$\giv b_1$};
    \node (s) [right =of b1] {$S$};
    \node (a2) [right =of s] {$A_2$};
    \node (b2) [right =0 of a2] {$\giv b_2$};
    \node (y) [right =of b2] {$Y^{b_1,b_2}$};

    \path (a1) edge[bend right=45] (s);
    \path (s) edge (a2);
    \path (b2) edge (y);
    \path (b1) edge[bend left=30] (y);
    \path (a1) edge[bend right=45] (a2);
  \end{tikzpicture}
  \caption{
    Time-dependent exposure $b$-SWIG constructed sequentially starting from latest exposure.
    Although there is a causal path from $A_1$ to $Y$ through $S$, there is no causal path from $A_1$ to $Y^{b_2}$ through $S$ due to the node splitting of $B_2$.
    Thus $S$ is not a descendant of $b_1$, and the distribution of $Y^{b_1,b_2}$ is identifiable in $\set{S}$ because $Y^{b_1,b_2} \ind S \giv (A_1, A_2)$.
  }
  \label{fig:time-dependent-exposure}
\end{figure}
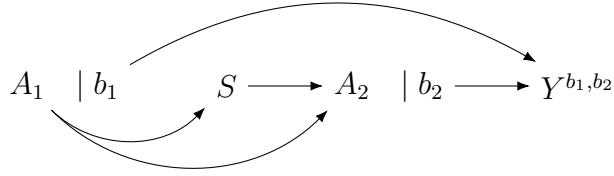

Multiple or time-dependent exposures can be handled by applying the exposure decomposition and node-splitting graph transformations to the exposures in reverse topological order.
In general, for any exposure $A_j$, the exposure decomposition into to $B_j$ and $C_j$ and the node-splitting operation introducing $b_j$ are applied before either operation is applied to an exposure $A_i$ that is an ancestor of $A_j$.
In the case of a single time-dependent exposure, reverse topological order is reverse chronological order.

Consider a randomized experiment evaluating protection from infectious respiratory illnesses by air purifiers ($A_2$) in classrooms of different sizes ($A_1$).
Enrollment ($S$) and randomization are stratified by room size, and suppose for simplicity that all enrolled classrooms operate at full capacity throughout school hours.
Incidence of infectious respiratory illnesses ($Y$) is measured only in the selected sample, and the following arguments apply regardless of whether room size and purification are measured for all classrooms or only those selected.
Figure~\ref{fig:time-dependent-exposure} depicts the final SWIG $\col{G}_{\textrm{ex}}^{b_1,b_2}$.
Beginning with the original DAG $\col{G}$, we first construct $\col{G}_{\textrm{ex}}^{b_2}$ by decomposing $A_2$ into $B_2$ (affecting $Y$) and $C_2$ (affecting nothing and thus excluded), then splitting $B_2$ into $B_2$ and $b_2$ to create the SWIG containing $Y^{b_2}$ as a node.
We then decompose $A_1$ into $B_1$ (affecting $Y^{b_2}$) and $C_1$ (affecting $S$ and $A_2$), then split $B_1$ into $B_1$ and $b_1$ to create the final SWIG $\col{G}_{\textrm{ex}}^{b_1, b_2}$ containing $Y^{b_1,b_2}$ as a node.
Due to exchangeability, the selection independence $S \ind Y^{b_1,b_2} \giv A_1, A_2$, and consistency we can identify
\begin{equation}
  \begin{aligned}
    \E\left[ Y^{b_1,b_2} \right]
    &= \E\left[ Y^{b_1,b_2} \mgiv A_1=b_1, A_2=b_2 \right], \\
    &= \E\left[ Y^{b_1,b_2} \mgiv A_1=b_1, A_2=b_2, S=1 \right], \\
    &= \E\left[ Y \mgiv A_1=b_1, A_2=b_2, S=1 \right].
  \end{aligned}
\end{equation}

Applying the graph operations to the exposures in reverse topological order is critical.
Because there is a causal path from $A_1$ to $Y$ through $S$, applying the exposure decomposition and node-splitting operations to $A_1$ before $A_2$ would replace $S$ with $S^{b_1}$ and thus prevent identification.
Indeed, $\E[Y^{b_1}]$ is not identifiable because the separable cohort condition is not met for classroom size ($A_1$) and selection ($S$).
In contrast, the separable cohort condition is met for air purification ($A_2$) and selection conditional on classroom size, thus $\E[Y^{b_2} \giv A_1]$ is identifiable.
If the distribution of clasroom sizes in the full population is known, $\E[Y^{b_2}]$ is also identifiable.

\subsection{Effect of selection in repeated COVID screening}
\label{sec:covid}

\begin{figure}
  \centering
  \begin{tikzpicture}[scale=1]
    \node (y1) at (0, 0) {$Y_1$};
    \node (y2) [right =3 of y1] {$Y_2^b$};
    \node (s1) [below =2 of y1] {$S_1$};
    \node (b) [right =0 of s1] {$\giv b$};
    \node (s2) [below =2 of y2] {$S_2$};

    \path (y1) edge (y2);
    \path (s1) edge[bend right=45] (s2);
    \path (b) edge (y2);
    \path (y1) edge[dashed] node[above left=0.2 and 1] {\small $b=1$} (s2);
  \end{tikzpicture}
  \caption{
    The dashed $Y_1 \rightarrow S_2$ is present only for $b=1$.
  }
  \label{fig:covid}
\end{figure}
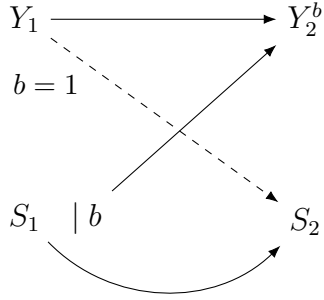

In longitudinal surveillance and isolation programs for infectious diseases in which the probability of testing a given individual on a given day depends on the time and result of their last test, the test-positive rate is a biased estimate of prevalence \citep{schnell2023overcoming}.
One determinant of the set of unbiased alternative estimators is whether testing causally affects the time until infection.
Here we present a simplified scenario in which such an effect is identifiable.

Let $Y_t=1$ indicate that an individual is infectious at time $t = 1, 2$, and $S_t$ indicate testing at time $t$ such that that if $S_t=1$ then $Y_t$ is observed for that individual.
Testing at $t=1$ is via simple random sampling.
Testing at $t=2$ depends on previous testing and, only if previously tested, the test result and thus also on infectiousness $Y_1$.
Figure~\ref{fig:covid} depicts the $b$-SWIG for analyzing the effect of testing at $t=1$ on infection by $t=2$.
Note that in the underlying DAG, $Y_1$ is a common cause of $S_2$ and $Y_2$, thus the simple test-positive rate is a biased estimate of prevalence: $\P[Y_2=1 \giv S_2=1] \neq \P[Y_2=1]$.

Separate applications of the separable cohort condition for each potential outcome yields identifiability: $\P[Y_2^0=1]$ because the separable cohort condition holds for $S_2$ unconditionally when $b=0$, and $\P[Y_2^1=1]$ because it holds given $Y_1$ when $b=1$, in which case $Y_1$ is observed.
Explicitly:
\begin{align}
  \P\left[Y_2^0=1\right]
  &= \P\left[ Y_2^0=1 \mgiv S_2=1, S_1=0 \right], \\
  &= \P\left[ Y_2=1 \mgiv S_2=1, S_1=0 \right], \label{eq:covid-consistency-0}
\end{align}
and
\begin{align}
  \P\left[Y_2^1=1\right]
  &= \sum_y \P\left[ Y_2^1=1 \mgiv Y_1=y \right] \P\left[Y_1=y\right], \label{eq:covid-total-prob} \\
  &= \sum_y \P\left[ Y_2^1=1 \mgiv S_2=1, Y_1=y, S_1=1 \right] \P\left[Y_1=y \right], \\
  &= \sum_y \P\left[ Y_2^1=1 \mgiv S_2=1, Y_1=y, S_1=1 \right] \P\left[Y_1=y \mgiv S_1=1 \right], \\
  &= \sum_y \P\left[ Y_2=1 \mgiv S_2=1, Y_1=y, S_1=1 \right] \P\left[Y_1=y \mgiv S_1=1 \right], \label{eq:covid-consistency-1}
\end{align}
where \eqref{eq:covid-total-prob} is due to the law of total probability, \eqref{eq:covid-consistency-0} and \eqref{eq:covid-consistency-1} to consistency, and the rest due to the independencies $[Y_2^0 \ind (S_1, S_2)]$, $[Y_2^0 \ind (S_1, S_2)]$, and $[Y_1 \ind S_1]$, which are all readable from the $b$-SWIG in Figure~\ref{fig:covid}.

\section{Conclusion}

Using the concept of separability \citep{robins2022interventionist, stensrud2022separable, stensrud2023conditional}, we have developed an alternative formulation of the potential outcomes approach to selection bias of \cite{kenah2023potential}.
As illustrated by examples, this new formulation enables nonparametric analysis of selection bias even in the case of multiple or time-dependent exposures, complex relationships between samples in which different sets of variables are observed, or causal effects of selection on the outcome.
Though not shown, it is also straightforward to deduce from the separable cohort and case-control conditions the results of \cite{mathur2024simple} for identifiability of the average treatment effect in the selected population, and the distinction of \cite{sjolander2023selection} between identifiability under outcome-associated selection and outcome-influenced selection.
The approach is flexible and multiple paths may be available to establishing identifiability of causal effects.
The overall strategy is straightforward:
\begin{enumerate}
\item Beginning with the DAG representing the causal system, add selection variables indicating membership into each sample;
\item Construct the $b$-SWIG such that totality is satisfied, applying graph transformations to multiple exposures in reverse topological order;
\item Optionally evaluate the cohort and case-control isolation conditions to narrow the space of possibly identifiable estimands;
\item Attempt to express the quantity of interest in terms satisfying the conditions of Theorem~\ref{thm:identification-between-samples}.
\end{enumerate}
Only the final step may require cleverness, as it is not always obvious whether or how such an expression can be constructed, such as in our example of identifying causal risks from a case-cohort study.
However, the separable cohort and case-control conditions, or the g-formula, may often be useful starting points, and successful establishment of identifiability by this approach constructs a template for a plug-in estimator.
Development of additional shortcuts and heuristics, as well as formal derivations of sequential versions of these cohort and case-control conditions, would be useful avenues for future work.

\bigskip
\begin{center}
{\large\bf SUPPLEMENTARY MATERIAL}
\end{center}

\begin{description}

\item[Data and code:] Not applicable.
  
\item[Appendix:] The appendix contains proofs of results described in the main text.

\end{description}

\iffalse
\bigskip
\begin{center}
{\large\bf ACKNOWLEDGMENTS}
\end{center}
\fi

\bibliographystyle{apalike}
\bibliography{article.bib}

\begin{thebibliography}{}

\bibitem[Avin et~al., 2005]{avin2005identifiability}
Avin, C., Shpitser, I., and Pearl, J. (2005).
\newblock Identifiability of path-specific effects.
\newblock In {\em IJCAI International Joint Conference on Artificial
  Intelligence}, pages 357--363.

\bibitem[Bareinboim and Pearl, 2013]{bareinboim2013general}
Bareinboim, E. and Pearl, J. (2013).
\newblock A general algorithm for deciding transportability of experimental
  results.
\newblock {\em Journal of Causal Inference}, 1(1):107--134.

\bibitem[Dahabreh and Hern{\'a}n, 2019]{dahabreh2019extending}
Dahabreh, I.~J. and Hern{\'a}n, M.~A. (2019).
\newblock Extending inferences from a randomized trial to a target population.
\newblock {\em European Journal of Epidemiology}, 34:719--722.

\bibitem[Greenland, 1977]{greenland1977response}
Greenland, S. (1977).
\newblock Response and follow-up bias in cohort studies.
\newblock {\em American Journal of Epidemiology}, 106(3):184--187.

\bibitem[Greenland et~al., 1999]{greenland1999causal}
Greenland, S., Pearl, J., and Robins, J.~M. (1999).
\newblock Causal diagrams for epidemiologic research.
\newblock {\em Epidemiology}, 10(1):37--48.

\bibitem[Hern{\'a}n et~al., 2001]{hernan2001marginal}
Hern{\'a}n, M.~A., Brumback, B., and Robins, J.~M. (2001).
\newblock Marginal structural models to estimate the joint causal effect of
  nonrandomized treatments.
\newblock {\em Journal of the American Statistical Association},
  96(454):440--448.

\bibitem[Hern{\'a}n et~al., 2004]{hernan2004structural}
Hern{\'a}n, M.~A., Hern{\'a}ndez-D{\'\i}az, S., and Robins, J.~M. (2004).
\newblock A structural approach to selection bias.
\newblock {\em Epidemiology}, 15(5):615--625.

\bibitem[Kenah, 2023]{kenah2023potential}
Kenah, E. (2023).
\newblock A potential outcomes approach to selection bias.
\newblock {\em Epidemiology}, 34(6):865--872.

\bibitem[Lu et~al., 2022]{lu2022toward}
Lu, H., Cole, S.~R., Howe, C.~J., and Westreich, D. (2022).
\newblock Toward a clearer definition of selection bias when estimating causal
  effects.
\newblock {\em Epidemiology}, 33(5):699--706.

\bibitem[Mathur and Shpitser, 2024]{mathur2024simple}
Mathur, M.~B. and Shpitser, I. (2024).
\newblock Simple graphical rules to assess selection bias in general-population
  and selected-sample treatment effects.
\newblock {\em American Journal of Epidemiology}.

\bibitem[Mohan and Pearl, 2021]{mohan2021graphical}
Mohan, K. and Pearl, J. (2021).
\newblock Graphical models for processing missing data.
\newblock {\em Journal of the American Statistical Association},
  116(534):1023--1037.

\bibitem[Prentice, 1986]{prentice1986case}
Prentice, R.~L. (1986).
\newblock A case-cohort design for epidemiologic cohort studies and disease
  prevention trials.
\newblock {\em Biometrika}, 73(1):1--11.

\bibitem[Richardson and Robins, 2013]{richardson2013single}
Richardson, T.~S. and Robins, J.~M. (2013).
\newblock Single world intervention graphs (swigs): A unification of the
  counterfactual and graphical approaches to causality.
\newblock {\em Center for the Statistics and the Social Sciences, University of
  Washington Series. Working Paper}, 128(30):2013.

\bibitem[Robins, 1986]{robins1986new}
Robins, J. (1986).
\newblock A new approach to causal inference in mortality studies with a
  sustained exposure period—application to control of the healthy worker
  survivor effect.
\newblock {\em Mathematical modelling}, 7(9-12):1393--1512.

\bibitem[Robins et~al., 2000]{robins2000marginal}
Robins, J.~M., Hernan, M.~A., and Brumback, B. (2000).
\newblock Marginal structural models and causal inference in epidemiology.
\newblock {\em Epidemiology}, 11(5):550--560.

\bibitem[Robins et~al., 2022]{robins2022interventionist}
Robins, J.~M., Richardson, T.~S., and Shpitser, I. (2022).
\newblock An interventionist approach to mediation analysis.
\newblock In {\em Probabilistic and Causal Inference: The Works of Judea
  Pearl}, pages 713--764. ACM.

\bibitem[Schnell et~al., 2023]{schnell2023overcoming}
Schnell, P.~M., Wascher, M., and Rempala, G.~A. (2023).
\newblock Overcoming repeated testing schedule bias in estimates of disease
  prevalence.
\newblock {\em Journal of the American Statistical Association},
  119(545):1--13.

\bibitem[Shpitser and Pearl, 2007]{shpitser2007counterfactuals}
Shpitser, I. and Pearl, J. (2007).
\newblock What counterfactuals can be tested.
\newblock {\em Proceedings of the 23rd Annual Conference on Uncertainty in
  Artificial Intelligence}, pages 437--444.

\bibitem[Shpitser et~al., 2022]{shpitser2022multivariate}
Shpitser, I., Richardson, T.~S., and Robins, J.~M. (2022).
\newblock Multivariate counterfactual systems and causal graphical models.
\newblock In {\em Probabilistic and Causal Inference: The Works of Judea
  Pearl}, pages 813--852. ACM.

\bibitem[Sj{\"o}lander, 2023]{sjolander2023selection}
Sj{\"o}lander, A. (2023).
\newblock Selection bias with outcome-dependent sampling.
\newblock {\em Epidemiology}, 34(2):186--191.

\bibitem[Stensrud et~al., 2023]{stensrud2023conditional}
Stensrud, M.~J., Robins, J.~M., Sarvet, A., Tchetgen~Tchetgen, E.~J., and
  Young, J.~G. (2023).
\newblock Conditional separable effects.
\newblock {\em Journal of the American Statistical Association},
  118(544):2671--2683.

\bibitem[Stensrud et~al., 2022]{stensrud2022separable}
Stensrud, M.~J., Young, J.~G., Didelez, V., Robins, J.~M., and Hern{\'a}n,
  M.~A. (2022).
\newblock Separable effects for causal inference in the presence of competing
  events.
\newblock {\em Journal of the American Statistical Association},
  117(537):175--183.

\end{thebibliography}

\clearpage

\renewcommand{\theequation}{A.\arabic{equation}}
\renewcommand{\thesection}{A.\arabic{section}}
\renewcommand{\thefigure}{A.\arabic{figure}}

\setcounter{equation}{0}
\setcounter{section}{0}
\setcounter{theorem}{0}
\setcounter{figure}{0}

\section{Relationship between $a$-SWIGs and $b$-SWIGS}
\label{app:a-swig-b-swig}

On an $a$-SWIG, d-separation implies conditional independence.
For example, if $\vec{K}$ d-separates $A$ from $Y^a$ on $\col{G}^a$, then $Y^a \ind A \giv \vec{K}$.
Similarly, when $\col{G}_{\textrm{ex}}^b$ represents intervention on the exposure component $B$, if $\vec{K}$ d-separates $B$ from $Y^b$ on $\col{G}_{\textrm{ex}}^b$, then $Y^b \ind B \giv \vec{K}$.
However, the deterministic relationship $A \equiv B \equiv C$ in the observed data means that we must be careful when conditioning on $A$ or $C$ because $Y^b \ind B \giv A$ is vacuously true yet insufficient for the identification of $\P[Y^b=y]$.
In fact, exchangeability is more usefully expressed as $Y^b \ind A \giv K$, and consistency as $A=b \implies Y=Y^b$.
The process for identification of quantities involving $Y^a$ then proceeds as follows:

\begin{enumerate}
\item Show that some quantity is identifiable under intervention on $B$, potentially using totality to equate $Y^b$ to $Y^a$.
\item Ensure that the identification formula only involves quantities observable under the constraint $A=B=C$.
\end{enumerate}

Expression of the $b$-SWIG in the abbreviated form used in this work helps to prevent errors arising in the second step by removing $B$ and $C$ from the graphs, and using $\col{G}_{\textrm{ex}}^{b}$ instead of $\col{G}_{\textrm{ex}}^{bc}$ prevents us from dealing with counterfactuals with $b\neq c$.
Below, we provide formal details of the relationship between $a$-SWIGs and $b$-SWIGs under two common causal models.

Let $\col{P}$ be a finest fully randomized causally interpretable structured tree graph (FFRCISTG) compatible with the DAG $\col{G}$ \citep{robins1986new, shpitser2022multivariate}.
That is, $\col{P}$ is generated by the one-step-ahead potential outcomes $V_i^{\vec{v}_{\pa_i}}$, the value of $V_i$ that would result from intervening to set the value of its parents (with indices $\pa_i$) in $\col{G}$ to $\vec{v}_{\pa_i}$.
Other counterfactuals are defined via recursive substitution.
For an index set $\vec{a}$, the value of $V_i$ that would result from intervening to set $\vec{V}_{\vec{a}} = \vec{v}_{\vec{a}}$ is
\begin{equation}
  V_i^{\vec{a}} = \left\{
    \begin{array}{ll}
      V_i^{\vec{v}_{\pa_i}}, & \textrm{if } \vec{a} = \pa_i, \\
      V_i^{\vec{v}_{\pa_i}, \vec{V}_{\pa_i \setminus \vec{a}}^{\vec{v}_{\vec{a}}}}, & \textrm{otherwise.}
    \end{array}
  \right.
\end{equation}
The SWIG $\col{G}^a$ is derived from the SWIG for intervening on every variable, $\col{G}^{\vec{v}}$, which implies the mutual independence of one-step-ahead potential outcomes:
\begin{equation}
  V_1 \ind V_2^{\vec{v}_{\pa_2}} \ind \cdots \ind V_N^{\vec{v}_{\pa_N}}.
\end{equation}

Now let $\col{Q}$ be a FFRCISTG compatible with the expanded DAG $\col{G}_{\textrm{ex}}$, with the additional determinism assumption $B^a \equiv C^a \equiv a$, which implies $A \equiv B \equiv C$ via consistency.
By construction of $\col{G}_{\textrm{ex}}$, each variable can have at most one of $B$ or $C$ as a parent (though variables can have both as ancestors).
Supposing $V_i$ is a child of $B$, the recursive substitution formula yields $V_i^{a=\beta} = V_i^{b=B^{a=\beta}} = V_i^{b=\beta}$, and similarly $V_i^{a=\beta} = V_i^{c=\beta}$ if $V_i$ is a child of $C$.
Thus the one-step-ahead potential outcomes in $\col{P}$ for variables that are children of $A$ on $\col{G}$ retain their interpretation in $\col{Q}$, as do all potential outcomes under the intervention setting $A=a$ by the recursive substitution formula.
However, positing that $B$ and $C$ exist as components of $A$ on which we could intervene separately means that the FFRCISTG associated with the expanded graph assumes independence between sets of potential outcomes that are descendants of one but not the other.
That is, if $\vec{W}$ consists of descendants of $B$ but not of $C$, and $\vec{X}$ consists of descendants of $C$ but not of $B$, then we assume $\vec{W}^{b} \ind \vec{X}^{c}$, and due to determinism and recursive substitution, $\vec{W}^{a=1} \ind \vec{X}^{a=0}$ and $\vec{W}^{a=0} \ind \vec{X}^{a=1}$.
These assumptions (and possibly many more) are made by the non-parametric structural equation model with independent errors (NPSEM-IE) for both the original and expanded graphs.
Two concrete examples are given below, along with a discussion of some implications.

\begin{figure}
  \centering
  \begin{subfigure}[t]{0.4\textwidth}
    \begin{tikzpicture}
      \node (A) at (0,0) {$A$};
      \node (a) [right =0 of A] {$\giv \;\, a$};
      \node (Ya) [above right =0.5 and 1 of a] {$Y^a$};
      \node (Sa) [below right =0.5 and 1 of a] {$S^a$};

      \path (a) edge (Ya);
      \path (a) edge (Sa);
    \end{tikzpicture}
  \end{subfigure} %
  ~\quad
  \begin{subfigure}[t]{0.4\textwidth}
    \begin{tikzpicture}
      \node (A) at (0,0) {$A$};
      \node (a) [right =0 of A] {$\giv \;\, a$};
      \node (Ba) [above right =0.5 and 1 of a] {$B^a$};
      \node (Ca) [below right =0.5 and 1 of a] {$C^a$};
      \node (b) [right =0 of Ba] {$\giv \;\, b$};
      \node (c) [right =0 of Ca] {$\giv \;\, c$};
      \node (Yb) [right =of b] {$Y^b$};
      \node (Sc) [right =of c] {$S^c$};

      \path (a) edge (Ba);
      \path (a) edge (Ca);
      \path (b) edge (Yb);
      \path (c) edge (Sc);
    \end{tikzpicture}
  \end{subfigure}
  \caption{
    (Left) A single world intervention graph (SWIG) $\col{G}^a$ representing intervention on $A$.
    (Right) A SWIG $\col{G}_{\text{ex}}^{abc}$ representing the possibility of intervening also on components $B$ and $C$ of $A$ separately.
  }
  \label{fig:swig-full-pswig}
\end{figure}
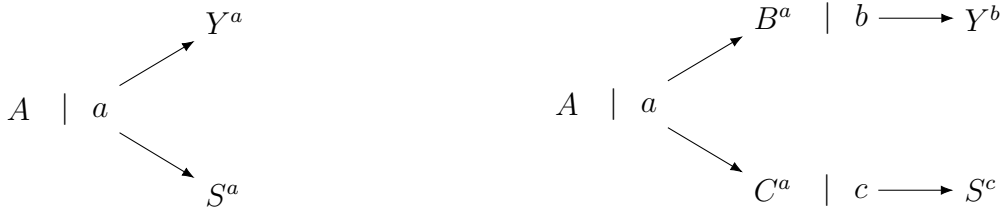

Figure~\ref{fig:swig-full-pswig} depicts an example of an $a$-SWIG $\col{G}^a$ for intervening on exposure $A$, and an expanded SWIG $\col{G}_{\text{ex}}^{abc}$ corresponding to intervening also on the components $B$ and $C$ of $A$ separately.
The FFRCISTG associated with $\col{G}^a$ makes the following independence assumptions:
\begin{equation}
  \begin{aligned}
    A &\ind Y^{a=0} \ind S^{a=0}, \\
    A &\ind Y^{a=1} \ind S^{a=1},
  \end{aligned}
\end{equation}
so that the variables on each line are mutually independent.
The FFRCISTG associated with $\col{G}_{\text{ex}}^{abc}$ makes the larger set of assumptions:
\begin{equation}
  A \ind B^{a} \ind Y^{b} \ind C^{a} \ind S^{c} \text{ for } a, b, c \in \{0,1\}.
\end{equation}
Under the determinism restriction $a \equiv B^a \equiv C^a$, the above assumptions become
\begin{equation}
  \begin{aligned}
    A \ind Y^{b=0} \ind S^{c=0}, \\
    A \ind Y^{b=0} \ind S^{c=1}, \\
    A \ind Y^{b=1} \ind S^{c=0}, \\
    A \ind Y^{b=1} \ind S^{c=1},
  \end{aligned}
\end{equation}
and further by recursive substitution the independence assumptions among variables in $\col{G}^a$ (i.e., in terms of intervention on $A$) become
\begin{equation}
  \begin{aligned}
    A \ind Y^{a=0} \ind S^{a=0}, \\
    A \ind Y^{a=0} \ind S^{a=1}, \\
    A \ind Y^{a=1} \ind S^{a=0}, \\
    A \ind Y^{a=1} \ind S^{a=1}.
  \end{aligned}
\end{equation}
Note that the first and last assumptions are the same as those made by the FFRCISTG associated with $\col{G}^a$.
The second and third assumptions include additional independencies $Y^{a=0} \ind S^{a=1}$ and $Y^{a=1} \ind S^{a=0}$ that are cross-world if we cannot define and intervene on the components $B$ and $C$ separately, but are not cross-world if we can do so.

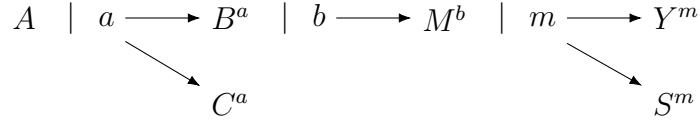
\begin{figure}
  \centering
  \begin{tikzpicture}
    \node (A) at (0,0) {$A$};
    \node (a) [right =0 of A] {$\giv \;\, a$};
    \node (Ba) [right =of a] {$B^a$};
    \node (b) [right =0 of Ba] {$\giv \;\, b$};
    \node (Mb) [right =of b] {$M^b$};
    \node (m) [right =0 of Mb] {$\giv \;\, m$};
    \node (Ym) [right =of m] {$Y^m$};
    \node (Sm) [below right =0.5 and 1 of m] {$S^m$};
    \node (Ca) [below right =0.5 and 1 of a] {$C^a$};

    \path (a) edge (Ba);
    \path (b) edge (Mb);
    \path (m) edge (Ym);
    \path (m) edge (Sm);
    \path (a) edge (Ca);
  \end{tikzpicture}
  \caption{
    A fully-intervened expanded SWIG $\col{G}_{\text{ex}}^{abm}$ for which the associated FFRCISTGs agrees with the FFRCISTG associated with the unexpanded SWIG $\col{G}^{am}$ when $a \equiv B^a \equiv C^a$ but not the NPSEM-IE associated with the underlying unexpanded DAG $\col{G}$.
  }
  \label{fig:non-separable-npsem}
\end{figure}

In the above example, the independencies among $A$ and potential outcomes induced by intervention on it are a strict superset of those assumed by the FFRCISTG associated with $\col{G}^a$, and exactly those assumed by the NPSEM-IE associated with the underlying DAG $\col{G}$.
However, neither is the case in general.
Figure~\ref{fig:non-separable-npsem} depicts a case where neither relationship holds.
Because $C$ has no descendants in $\col{G}_{\text{ex}}$, the FFRCISTG associated with $\col{G}_{\text{ex}}^{abm}$ does not make any additional assumptions about variables induced by intervention on variables in $\col{G}^{am}$ that the FFRCISTG associated with $\col{G}^{am}$ does not make.
However, the NPSEM-IE associated with $\col{G}$ makes the assumptions $Y^{a=0} \ind S^{a=1}$ and $Y^{a=1} \ind Y^{a=0}$.
In contrast to the \emph{possibly} cross-world assumptions made by the FFRCISTG associated with the expanded SWIG in the previous example, these assumptions are \emph{necessarily} cross-world in the sense that, even if one could intervene on any components of $A$, and even on any components of $M$ separately, one still could not observe $Y^{a=1}$ and $S^{a=0}$ together (nor $Y^{a=0}$ and $S^{a=1}$ together).
The salient property of the system depicted is that $M$ is a common cause of $Y$ and $S$ that is also causally affected by $A$.
It is sometimes said that to observe both $Y^{a=0}$ and $S^{a=1}$, $M$ would need to be a \emph{recanting witness} \citep{avin2005identifiability} in that it would need to take the value $M^{a=0}$ to observe $Y^{a=0}$ and $M^{a=1}$ to observe $S^{a=1}$.

\section{Proofs of sufficiency of separable cohort and case-control conditions}

Suppose the following conditions hold:
\begin{enumerate}
\item \textbf{Totality:} $Y^b=Y^a$;
\item \textbf{Cohort isolation:} $S^b=S$;
\item \textbf{Separable cohort condition:} $(S^b, A) \ind Y^b \giv \vec{K}$.
\end{enumerate}
By totality and cohort isolation, the separable cohort condition may be written $(S, A) \ind Y^a \giv \vec{K}$.
Then $\P[Y^a=y] = \P[Y^a=y \giv A=a, S=1] = \P[Y=y \giv A=a, S=1]$ by the separable cohort condition and consistency.
Alternatively, working entirely from the $b$-SWIG,
\begin{equation}
  \begin{aligned}
    \P[Y=y \giv A=b, \vec{K}=\vec{k}, S=1]
    &= \P[Y^b=y \giv A=b, \vec{K}=\vec{k}, S=1], \\
    &= \P[Y^b=y \giv A=b, \vec{K}=\vec{k}], \\
    &= \P[Y^b=y \giv \vec{K}=\vec{k}].
  \end{aligned}
\end{equation}
By totality, $Y^b=Y^a$.
If $\vec{K}$ can be decomposed into $(\vec{X}, \vec{W})$ such that $Y^b \ind \vec{W} \giv \vec{X}$, then the potential outcome distribution in the final line can be equated to one conditioning only on $\vec{X}$ instead of $\vec{K}$.
If the decomposition satisfies $Y^b \ind A \giv \vec{X}$ and $Y^b \ind \vec{W} \giv (A, \vec{X})$ then the substitution of $\vec{X}$ for $\vec{K}$ may be made at the penultimate line and carried through.

Now suppose instead that the following conditions hold:
\begin{enumerate}
\item \textbf{Totality:} $Y^b=Y^a$;
\item \textbf{Case-control isolation:} All causal paths from $B$ to $S$ pass through $Y$;
\item \textbf{Separable case-control condition:} $(S^b, Y^b) \ind A \giv \vec{K}$.
\end{enumerate}
By Lemma~\ref{lem:observable-conditioning} (which assumes case-control isolation), Bayes' Theorem, consistency, and exchangeability (implied by the separable case-control condition),
\begin{equation}
  \begin{aligned}
    &\hspace{-1em}\frac{
      \P[A=1 \giv Y=1, K=k, S=1] / \P[A=0 \giv Y=1, \vec{K}=\vec{k}, S=1]
    }{
      \P[A=1 \giv Y=0, K=k, S=1] / \P[A=0 \giv Y=0, \vec{K}=\vec{k}, S=1]
    } \\
    &=
    \frac{
      \P[A=1 \giv Y=1, K=k] / \P[A=0 \giv Y=1, \vec{K}=\vec{k}]
    }{
      \P[A=1 \giv Y=0, K=k] / \P[A=0 \giv Y=0, \vec{K}=\vec{k}]
    } \\
    &=
    \frac{
      \P[Y=1 \giv A=1, K=k] / \P[Y=0 \giv A=1, \vec{K}=\vec{k}]
    }{
      \P[Y=1 \giv A=0, K=k] / \P[Y=0 \giv A=0, \vec{K}=\vec{k}]
    }, \\
    &=
    \frac{
      \P[Y^1=1 \giv A=1, K=k] / \P[Y^1=0 \giv A=1, \vec{K}=\vec{k}]
    }{
      \P[Y^0=1 \giv A=0, K=k] / \P[Y^0=0 \giv A=0, \vec{K}=\vec{k}]
    }, \\
    &=
    \frac{
      \P[Y^1=1 \giv K=k] / \P[Y^1=0 \giv \vec{K}=\vec{k}]
    }{
      \P[Y^0=1 \giv K=k] / \P[Y^0=0 \giv \vec{K}=\vec{k}]
    }.
  \end{aligned}
\end{equation}
By totality, $Y^1$ equivalently represents $Y^{b=1}$ or $Y^{a=1}$, and similarly for $Y^0$.

If $\vec{K}$ can be decomposed into $(\vec{X}, \vec{W})$ with $Y^b \ind W \giv \vec{X}$, then the causal odds ratio conditional on $\vec{K}$ in the final line above may be further equated to the causal odds ratio conditional on $\vec{X}$ only.
If the decomposition satisfies $Y^b \ind A \giv \vec{X}$ and $Y^b \ind \vec{W} \giv (A, \vec{X})$ then the substitution may be made at the penultimate line and carried through.

\section{Equivalence of conditions with strict non-descent requirements}

\subsection{Cohort conditions}
\label{app:cohort-equivalence}

The analytic cohort condition requires $(S^a, A) \ind Y^a \giv \vec{X}$ with $\vec{X}$ containing no descendants of $A$.
The separable cohort condition requires $(S^b, A) \ind Y^b \giv \vec{K}$, where $\vec{K}$ contains no descendants of $b$ (whose descendants are a subset of those of $A$).
If we require $\vec{K}$ to contain no descendants of $A$, then to show that the conditions are equivalent, we must show that $(S^b, A) \ind Y^b \giv \vec{K}$ if and only if $(S^a, A) \ind Y^a \giv \vec{K}$.
We divide this task into establishing the equivalence of $A \ind Y^b \giv \vec{K}$ and $A \ind Y^a \giv \vec{K}$, then the equivalence of $S^b \ind Y^b \giv (A, \vec{K})$ and $S^a \ind Y^a \giv (A, \vec{K})$.

To show that $A \ind Y^b \giv \vec{K}$ if and only if $A \ind Y^a \giv \vec{K}$, note that, conditioning on $\vec{K}$, there are one-to-one correspondences among the following sets:
\begin{enumerate}
\item Open paths between $A$ and $Y^a$ on the $a$-SWIG $\col{G}^a$,
\item Open back-door paths from $Y$ into $A$ on the DAG $\col{G}$,
\item Open back-door paths from $Y$ into $A$ on the separated DAG $\col{G}_{\textrm{ex}}$,
\item Open back-door paths from $Y$ into $B$ on $\col{G}_{\textrm{ex}}$,
\item Open paths between $B$ and $Y^b$ on the $b$-SWIG $\col{G}_{\textrm{ex}}^b$.
\end{enumerate}
Thus the equivalence.

To show that $S \ind Y^b \giv (A, \vec{K})$ if and only if $S^a \ind Y^a \giv (A, \vec{K})$, note that, conditioning on $(A, \vec{K})$, there are one-to-one correspondences among the following sets:
\begin{enumerate}
\item Open paths between $S^a$ and $Y^a$ on the $a$-SWIG $\col{G}^a$,
\item Open paths between $S$ and $Y$ on the DAG $\col{G}$,
\item Open paths between $S$ and $Y$ on the separated DAG $\col{G}_{\textrm{ex}}$,
\item Open paths between $S$ and $Y^b$ on the $b$-SWIG $\col{G}_{\textrm{ex}}^b$.
\end{enumerate}
Thus the equivalence.

\subsection{Case-control conditions}
\label{app:case-control-equivalence}

The analytic case-control condition requires $(\vec{W}^a, A) \ind Y^a \giv \vec{X}$ with $\vec{X}$ containing no descendants of $A$ and $\vec{W}$ containing no descendants of $Y$ with $S^y \ind A \giv (Y, \vec{X}, \vec{W})$.
The separable case-control condition requires case-control isolation (all causal paths from $B$ to $S$ pass through $Y$) and $(Y^b, S^b) \ind A \giv \vec{K}$.

The two conditions imply different identifiability results: The analytic condition implies that the source population causal odds ratio given $\vec{X}$ is equal to the exposure odds ratio given $(\vec{X}, \vec{W})$ in the selected population, while the separable condition implies that the source population causal odds ratio given $\vec{K}$ is equal to the exposure odds ratio given $\vec{K}$ in the selected population.
As shown in the proof of sufficiency of the separable case-control condition, if we require $\vec{K}=(\vec{X}, \vec{W})$ with $Y^b \ind A \giv \vec{X}$ and $Y^b \ind \vec{W} \giv (A, \vec{X})$, then we get the result implied by the analytic case-control condition.
We wish to show that if we also require that $\vec{X}$ contains no descendants of $A$ and $\vec{W}$ contains no descendants of $Y$, it is equivalent to the analytic case-control condition.

First, note that $Y^b \ind A \giv \vec{X}$ and $Y^b \ind W \giv (A, \vec{X})$ if and only if $(\vec{W}, A) \ind Y^b \giv \vec{X}$, which is equivalent to the requirement $(\vec{W}^a, A) \ind Y^a \giv \vec{X}$ in the analytic case-control condition.
This equivalence follows from the same argument that establishes the equivalence of $(S^b, A) \ind Y^b \giv \vec{X}$ and $(S^a, A) \ind Y^a \giv \vec{X}$ in the cohort conditions (where $S^b = S$ due to cohort isolation).
It remains is to show that given the requirements of the previous paragraph, case-control isolation and $(Y^b, S^b) \ind A \giv (\vec{X}, \vec{W})$ together are equivalent to $S^y \ind A \giv (Y, \vec{X}, \vec{W})$.
The independence $(Y^b, S^b) \ind A \giv (\vec{X}, \vec{W})$ is equivalent to $S^b \ind A \giv (Y^b, \vec{X}, \vec{W})$ and $Y^b \ind A \giv (\vec{X}, \vec{W})$.
Because $(\vec{W}, A) \ind Y^b \giv \vec{X}$, the latter is already implied by both the analytic and separable case-control conditions.
Thus the only thing left to show that case-control isolation and $S^b \ind A \giv (Y^b, \vec{X}, \vec{W})$ together are equivalent to $S^y \ind A \giv (Y, \vec{X}, \vec{W})$.

We now show by contradiction that $S^y \ind A \giv (Y, \vec{X}, \vec{W})$ implies case-control isolation under previously-established requirements.
Suppose that case-control isolation does not hold.
Then there exists a causal path from $B$ to $S$ that does not go through $Y$ and therefore is not closed by conditioning on $Y$.
Since we have established that $Y^b \ind \vec{W} \giv (A, \vec{X})$ and $\vec{X}$ does not contain any descendants of $A$, $W$ cannot contain a descendant of $B$ and thus cannot close any causal paths from $B$ to $S$ (nor can $\vec{X}$).
Thus there is an open causal path between $B$ and $S$ (which implies one from $A$ to $S$) that does not pass through $Y$ and remains after the node-splitting operation that yields $\col{G}^y$.
But then $S^y \notind A \giv (Y, \vec{X}, \vec{W})$, which is a contradiction.
Therefore $S^y \ind A \giv (Y, \vec{X}, \vec{W})$ implies case-control isolation.

Finally, we show that $S^b \ind A \giv (Y^b, \vec{X}, \vec{W})$ implies $S^y \ind A \giv (Y, \vec{X}, \vec{W})$ under case-control isolation.
Note that there are one-to-one correspondences among the following sets (regardless of case-control isolation):
\begin{enumerate}
\item Open paths betwen $A$ and $S^y$ on the $y$-SWIG $\col{G}^y$, conditioning on $(Y, \vec{X}, \vec{W})$;
\item Open paths between $A$ and $S$ on DAG $\col{G}$, conditioning on $(Y, \vec{X}, \vec{W})$;,
\item Open paths between $A$ and $S$ on the separated DAG $\col{G}_{\textrm{ex}}$, conditioning on $(Y, \vec{X}, \vec{W})$.
\end{enumerate}
Under case-control isolation, the final set above has a one-to-one correspondence with open paths between $A$ and $S^b$ on $\col{G}_{\textrm{ex}}^b$ after conditioning on $(Y^b, \vec{X}, \vec{W})$, because any path from $A$ to $S$ broken by the node-splitting operation on $B$ must be a causal path not passing through $Y$ (which is prohibited by case-control isolation).
Thus the equivalence.

\section{Implication of isolation conditions by analytic conditions}

\subsection{The analytic cohort condition implies cohort isolation}
\label{app:cohort-isolation-necessary}

Suppose cohort isolation fails, i.e., there is a causal path from $B$ to $S$ on $\col{G}_{\textrm{ex}}$.
We will show that the analytic cohort condition also fails.
In particular, $S^a \notind Y^a \giv (A, \vec{X})$.
Because there is a causal path from $B$ to $Y$ on $\col{G}_{\textrm{ex}}$ by construction, the existence of a causal path from $B$ to $S$ on  $\col{G}_{\textrm{ex}}$ implies that---absent any conditioning---there is an open (possibly non-causal) path between $Y$ and $S$ that contains no non-descendants of $B$ (except possibly $B$ itself).
In particular, that path contains neither $A$ nor any variable in $\vec{X}$, so that the node-splitting operation yielding $\col{G}_{\textrm{ex}}^a$ does not sever the path nor does conditioning on $(A, \vec{X})$ close the path.
Thus $(A, \vec{X})$ does not d-separate $Y^a$ from $S^a$ on $\col{G}_{\textrm{ex}}^a$, (nor on $\col{G}^a$) so $S^a \notind Y^a \giv (A, \vec{X})$.

\subsection{The analytic case-control condition implies case-control isolation}
\label{app:case-control-isolation-necessary}

First, note that the expanded DAG $\col{G}_{\textrm{ex}}$ underlying the SWIGs in Figure~\ref{fig:separable-case-control} does not satisfy cohort isolation but does satisfy the analytic case-control condition.
Thus the analytic case-control condition does not imply cohort isolation.

Now suppose case-control isolation fails, so there is a causal path from $B$ to $S$ on $\col{G}_{\textrm{ex}}$ that does not pass through $Y$.
We will show that the analytic case-control condition then fails.
In particular, if $S^y \ind A \giv (Y, \vec{X}, \vec{W})$ then $\vec{W}^a \notind Y^a \giv (A, \vec{X})$.

The causal path from $B$ to $S$ on $\col{G}_{\textrm{ex}}$ that does not pass through $Y$ contains no non-descendants of $B$ other than $B$ itself.
It can be extended to a causal path from $A$ to $S$ on $\col{G}$ not passing through $Y$ that contains no non-descendants of $A$ other than $A$ itself.
Such an extended path remains open when conditioning on $(Y, \vec{X})$ where $\vec{X}$ contains no descendants of $A$, and it is not-severed by the node-splitting operation yielding $\col{G}^y$.
Thus $(Y, \vec{X})$ does not d-separate $S^y$ from $A$ on $\col{G}^y$.

However, suppose that $S^y \ind A \giv (Y, \vec{X}, \vec{W})$, so that $(Y, \vec{X}, \vec{W})$ does d-separate $S^y$ from $A$ on $\col{G}^y$.
Then a component of $W$ must lie on the causal path from $B$ to $S$ on $\col{G}_{\textrm{ex}}$ that does not pass through $Y$, so $\vec{W}^a \notind Y^a \giv (A, \vec{X})$.

\section{Proof of Lemma~\ref{lem:observable-conditioning}}
\label{app:observable-conditioning}

Suppose $S \ind A \giv (Y, \vec{K})$.
Then $(Y, \vec{K})$ d-separates $S$ from $A$ on $\col{G}$ and therefore on $\col{G}_{\textrm{ex}}$ as well.
Any open path between $S^b$ and $A$ on $\col{G}_{\textrm{ex}}^{b}$ must correspond to an open path between $S$ and $A$ on $\col{G}_{\textrm{ex}}$.
Because there are no such paths, we must have $S^{b} \ind A \giv (Y^b, \vec{K})$.

Now suppose $S^b \ind A \giv (Y^b, \vec{K})$.
By definition of $B$, we have $S^b \ind B \giv (Y^b, \vec{K})$, so that any open path between $b$ and $S^b$ on $\col{G}_{\textrm{ex}}^b$ must correspond to a causal (i.e., directed) path from $B$ to $S$ that does not pass through $Y$.
The case-control isolation condition rules out any such path.
Because there is no open path between $S^b$ and $b$ on $\col{G}_{\textrm{ex}}^b$, transformation to $\col{G}_{\textrm{ex}}$ does not yield any open path between $S$ and $B$.
Thus $S \ind A \giv (Y, \vec{K})$.

\end{document}